%% file: sf04.tex
\documentclass[twocolumn]{revtex4-2} 

\input{settings}
\newcommand{\resub}[1]{#1}
\begin{document}

\title{Finite-temperature superfluid depletion of disordered Bose gases}

\author{Cord A.\ M\"uller 
}
\affiliation
{Deutsche Akademie für Metrologie beim Bayerischen Landesamt für Maß und Gewicht, Wittelsbacher Str. 14, 83435 Bad Reichenhall, Germany}

\email{cord.mueller@lmg.bayern.de}

\begin{abstract} 
At zero temperature, homogeneous interacting Bose-condensed fluids are entirely superfluid, with remarkable transport properties. 
A non-superfluid, normal component is induced by finite temperatures and spatial inhomogeneity, the combined effects of which are rather intriguing, and difficult to describe quantitatively. 
By inhomogeneous Bogoliubov theory, applicable to weakly interacting condensed Bose gases in static external potentials with arbitrary spatial correlations, we calculate the normal fluid density via the transverse current-current correlation. 
We obtain finite-temperature disorder corrections to the normal fraction known since Laudau's seminal two-fluid theory, using diagrammatic perturbation theory for systems of any dimensionality, with closed analytical expressions to leading, quadratic order in disorder strength.  
\end{abstract} 

\maketitle

\section{Introduction} 

One of the most spectacular features of interacting Bose-Einstein condensed gases is their
ability to sustain superfluidity. This wholly nontrivial, macroscopic
phenomenon is central for understanding the peculiar transport properties 
of superfluid Helium-4 \cite{Glyde2018}, ultracold atomic gases \cite{Pitaevskii2003}, polariton
condensates \cite{Amo2009}, etc. 
An especially intriguing question in many realizations is the impact of spatial inhomogeneity on the superfluid
fraction, which is known to be depleted by static impurities or external potentials.

The total mass density of a superfluid, $\rho=\rhos+\rhon$, is
the sum of the \resub{properly} superfluid mass density $\rhos$ that can flow without
dissipation, and the normal density  $\rhon$ that is  
responsible for viscous effects. 
For a homogeneous system, Landau's two-fluid theory 
\cite{LandauStatPhys2} predicts on elementary 
grounds that the normal density is due to the presence of 
elementary excitations: 
\be\label{rhonLandau}
\rhon =\frac{1}{d} \int \frac{\rmd^dp}{(2\pi)^d}  p^2
(-\partial_{\ep{\bp}} \nu_{\ep{\bp}}).
\ee
Here, $\nu_{\ep{}}=[\exp\{\beta\ep{}\}-1]^{-1}$ is the
canonical distribution of bosonic excitations with
dispersion $\resub{\ep{}={}}\ep{\bp}$ 
\resub{as function of momentum $\bp$ expressed in units where $\hbar=1$}
at inverse temperature $\beta=1/\kB
T$.  
At zero temperature, there are no excitations, and the normal fraction $\fn=\rhon/\rho$
vanishes. 

When an external potential renders the system spatially inhomogeneous,
it modifies the spectrum of excitations, and one should expect
the temperature-dependent normal fraction \eqref{rhonLandau} to be affected as well. 
Indeed, already at zero temperature, external potentials are known to induce a normal
component. This normal component
depletes the superfluid and is thus accountable for dissipative
responses with, e.g., finite drag forces at arbitrarily low fluid
velocity, notably below Landau's critical velocity. 
However, the derivation of Laudau's classical result \eqref{rhonLandau} hinges on translation invariance in a homogeneous setting 
\cite{Pitaevskii2003}, and it is not evident if, and how, this relation can be modified to include effects of spatial inhomogeneity \cite{Abdullaev2012}. 

In this article, we present an analytical calculation of the normal density  
induced by an external potential in interacting Bose-condensed fluids, computed via the
microscopic transverse current response, valid at zero and finite
temperature. 
We take special care to discuss spatially correlated
potentials, which have lately become
under detailed scrutiny with ultracold atoms in optical
lattices and random potentials 
\cite{Lye2005,Fallani2007,Shapiro2012,Chauveau2023,Geier2025,Perez-Cruz2025,Rabec2026}. 
The correlation lengths of such potentials can be tuned to be of the order of, if not much longer than, the coherence length of the condensate. 
\resub{Therefore, it is inappropriate to model such potentials as spatially delta-correlated processes.} 
 
\resub{Superfluid cold atomic gases are studied in various effective dimensionalities $d=1,2,3$, depending on the geometry of the external (magneto-optical) trap. 
Frequently, these systems are 3D fluids from a microscopic point of view, with a mean-field interaction energy (chemical potential $\mu$) exceeding the characteristic trapping frequencies in all directions (as opposed to strict 2D or even 1D systems, which require a much tighter confinement along one or even two transverse directions). 
At sufficiently low temperatures, one can thus study (quasi-condensed) bosonic superfluids, even in lower dimensions $d=1, 2$, provided their phase-coherence length is larger than the system size.  
The elementary excitations then only explore effective dimensions lower than $d=3$, 
with little or even --  in optical box potentials \cite{Gaunt2013,Navon2021,Chauveau2023} -- no trapping at all in the soft longitudinal directions, up to the system size.  
}

We approach this setting using inhomogeneous Bogoliubov theory, applicable to weakly interacting, (quasi-)condensed Bose gases of any dimensionality. \resub{The dynamics of elementary low-energy excitations in presence of an external potential that breaks translation invariance is governed by} 
the effective, quadratic Hamiltonian derived in Ref.~\cite{Gaul2011a} whose main advantage is that it describes the excitations of the \resub{real,} deformed condensate (or inhomogeneous Bogoliubov vacuum) instead of a fictitious homogeneous condensate \cite{Huang1992,Giorgini1994,Kobayashi2002}. 
Our results extend well-established analytical approaches to superfluid depletion \cite{Lopatin2002,Yukalov2007,Yukalov2007a} in
two important aspects, namely by keeping track of the aforementioned spatial potential
correlations and by systematically including all contributions to a given order in the external one-body potential $V(\br)$.
\resub{By a perturbative expansion in powers of the random potential $V$, we provide an analytical expression to order $V^2$ of the finite-temperature normal density, or superfluid depletion, a quantity that is pivotal for describing the non-equilibrium transport properties of interacting condensed Bose gases in systems with broken translational invariance. 
This work thus } 
complements previous results on the finite-temperature condensate fraction \cite{Gaul2014} \resub{-- a quantity of equilibrium thermodynamics  --} and delivers on a promise made in a previous work on the Josephson relation for disordered superfluids \cite{Mueller2015}.

After setting the stage in Sec.~\ref{formalism.sec}, expressing the transverse current correlation in terms of the inhomogeneous Bogoliubov formalism, we
discuss the contributions of single- and pair-excitation type in 
two separate sections, \ref{g1.sec} and \ref{g2.sec}. Whereas the
single-excitation superfluid depletion reproduces a well-known, temperature-independent result, 
we succeed to obtain temperature-dependent disorder corrections to Landau's
relation \eqref{rhonLandau} via the pair-excitation response. 
Well-behaved analytical
expressions are obtained for the normal\resub{, non-superfluid} fraction to second order in the
external potential strength. Particularly simple results are found in the
Thomas-Fermi regime of smooth potentials, whose correlation length is
larger than the condensate healing length.

\section{Starting lineup} 
\label{formalism.sec}

We consider a dilute gas \resub{with average density $n$} of massive bosons \resub{($m>0$)}
interacting via short-range repulsion
(s-wave interaction parameter $g>0$) 
\resub{and average mean-field interaction energy (chemical potential) $\mu = gn$,} 
subject\resub{ed} to a stationary, external \resub{one-body} potential
$V(\br)$. 
The formalism developed \resub{here} suits any such potential; 
it is applied 
\resub{in the following} to a spatially ergodic random potential with zero mean 
$\avg{V(\br)} = L^{-d}\int\rmd \br V(\br) =0 $ (without loss of generality\resub{, since a finite mean merely redefines $\mu$}) 
and finite variance $\avg{V(\br)^2}=:V^2$. 
\resub{In cold-atom experiments, such potentials are typically realised via laser speckle fields, which are sufficiently far detuned from an  optical atomic resonance to induce an effective, conservative, spatially random one-body potential \cite{Kuhn2007}. 
The global trap potential $V_\text{trap}(\br)$ only enters implicitly by determining both the overall size $L$ of the system and the number of dimensions $d$ accessible to the low-energy (phononic) excitations of the system. 
The density in the clean case $V(\br) = 0 $ is taken constant over the system size $L$; we are not concerned with boundary or finite-size effects, as recently studied theoretically by Berger et al. \cite{Berger2026a,Berger2026b}.
In terms of the characteristic energies involved, we consider the regime $\mu \gg V \gg E_L = 1/(2mL^2) \to 0$ (in units such that $\hbar=1$), where the random potential $V(\br)$ has a noticeable effect on bulk transport properties, but does not fragment the condensate. 
A priori, we make no restricting assumptions about the thermal energy $\kB T$ and the random-potential correlation energy $E_\sigma=1/(2m\sigma^2)$ (with $\sigma$ the spatial correlation length, see eq.~\eqref{Cpsigma} below), apart from the fact that the temperature should be much lower than the BEC transition temperature. It is the central aim of this paper to assess the dependence of the normal component on the potential correlation length and temperature as freely variable parameters. 
}

The normal\resub{, non-superfluid} mass density \resub{can be defined} 
by the fluid's transverse response  
to the walls in the $xy$-plane of an open-ended   
vessel that are moving into the $z$-direction and thereby drag along only the normal component \cite{LandauStatPhys2,Baym1969}:   
\be\label{rhonggdef}
\rhon = L^d \int_0^\beta \rmd
\tau \avg{\xpct{\hat{g}_{z\bk_\perp}(\tau)\hat{g}_{z(-\bk_\perp)}(0)}}_{\bk_\perp\to0}. 
\ee
The brackets $\xpct{\cdot}$ denote an expectation value  
in the thermal equilibrium state of the fluid at rest at inverse temperature $\beta=1/\kB T$;  
we use 
Matsubara's imaginary-time representation  
$\hat A(\tau) = e^{\tau \hat H} \hat A e^{-\tau \hat H}$.
The vector 
\be\label{gdef}
 \hat\bg_\bk = \frac{1}{2iL^d} \intddr \,e^{-i\bk\cdot\br}
 \left [\hat\psi^\dagger(\br) \nabla \hat\psi(\br)-
\left(\nabla\hat\psi^\dagger(\br)\right)\hat\psi(\br)
\right]
\ee 
is the momentum (or mass current) density in units where
$\hbar=1$. 
The transverse part $\hat{g}_{z\bk_\perp}$ of the current's
$z$-component is selected by choosing $\bk_\perp=(k_x,k_y,0)$ in the $xy$-plane. 
(In contrast, the longitudinal current
response with $\bk$ along the $z$-axis would  
describe a vessel with closed ends in that direction, which carries along the
entire fluid, and thus yields the total density, as recalled in appendix \ref{numberconserving.sec}.)

At temperatures well below the Bose-Einstein condensation
temperature, the boson field operator $\hat\psi(\br) = \Phi(\br)
+\delta\hat\psi(\br)$ splits into the macroscopically populated mean-field condensate $\Phi(\br)$
and its excitations 
$\delta\hat\psi(\br)$. 
\resub{As recalled in appendix \ref{InhBogoTheory.sec},}
a saddle-point expansion of the full
Hamiltonian around the deformed condensate yields an effective, quadratic Hamiltonian for non-interacting Bogoliubov excitations \resub{ that can be cast} in
the \resub{standard impurity-scattering} form \cite{Gaul2011a} 
\be\label{InhBogH}
\hat H= \hat H_0 + \hat \calV .
\ee 
Here, the free Hamiltonian 
\be\label{H0}
\hat H_0 = \sum_\bk\ep{\bk}\gd{\bk}\g{\bk}
\ee
features the celebrated Bogoliubov dispersion
$\ep{\bk}=[\epn{\bk}(\epn{\bk}+2\mu)]^{1/2}$, with $\epn{\bk}=k^2/2m$ the free dispersion. 
$\g{\bk}$ ($\gd{\bk}$)
annihilates (creates) a Bogoliubov excitation, called ``bogolon'' for
short, obeying the canonical bosonic commutation relations
$[\g{\bk},\gd{\bk'}]= \delta_{\bk\bk'}$ and $[\g{\bk},\g{\bk'}]= 0 $.
The one-body potential  
\be \label{calVdef} 
\hat\calV = \frac{1}{2}\sum_{\bk\bk'} (\gd{\bk},\g{-\bk}) 
\calV_{\bk\bk'}
\cvect{\g{\bk'}}{\gd{-\bk'}}
\ee
describes elastic scattering of bogolons. 
Importantly, the effective potential  
$\calV
=\begin{smallpmatrix} W & 
  Y\\ Y & W\end{smallpmatrix}
$
encodes all the information about the underlying deformed 
condensate with density $\nc(\br) = |\Phi(\br)|^2$ and Fourier components 
${\nc}_{\bk} = L^{-d}\int\rmd^dr \,e^{-i\bk\cdot\br} \nc(\br)$, 
all of which are nonlinear functionals of the bare external potential $V(\br)$. 
\resub{The effective potential $\hat\calV$ describes how the excitations of the real, inhomogeneous Bogoliubov vacuum (namely, the deformed condensate) are scattered by the spatial inhomogeneities. 
It thus accounts for all effects of the external potential $V(\br)$,  both on the condensate and its excitations.}
With this approach, effects of
interaction are taken into account on the mean-field level, and
we can concentrate on the impact of spatial inhomogeneity on the non-interacting excitations by using standard
recipes of single-particle quantum transport theory.

In order to evaluate Eq.~\eqref{rhonggdef} within our approach, we must express the particle current \eqref{gdef} in terms of bogolons. 
We use the density-phase representation $\hat\psi(\br) =
e^{i\dph(\br)}[\nc(\br)+\dn(\br)]^{1/2}$, where the current develops as
$\hat\bg(\br) = \hat\bg^{[1]}(\br) + \hat\bg^{[2]}(\br)$ into the sum of a single- and pair-bogolon
contribution: 
\begin{eqnarray}
\hat\bg^{[1]}(\br)& = & \nc(\br)\nabla \dph(\br), \label{currentg1}\\
\hat\bg^{[2]}(\br) & = & \frac{1}{2}[\dn(\br) \nabla
\dph(\br) + (\nabla
\dph(\br))\dn(\br)].
\label{currentg2}
\end{eqnarray} 
There is no zeroth-order term because
the background condensate 
carries no current, and higher-order terms are
beyond the accuracy of the quadratic Hamiltonian used to
evaluate the normal density \eqref{rhonggdef}. 
 
In the Fourier-Bogoliubov representation 
\begin{align}
\dn_\bk &= a_\bk\sqrt{\nc}(\g{\bk}+\gd{-\bk})\\
\dph_\bk &=  (\g{\bk}-\gd{-\bk})/(2ia_\bk\sqrt{\nc})
\end{align}
where $a_\bk=(\epn{\bk}/\ep{\bk})^{1/2}$, 
the single-bogolon current \eqref{currentg1} becomes  
\be\label{g1k} 
\hat\bg^{[1]}_\bk = \frac{1}{2\sqrt{\Nc}} \sum_\bp \bp
\frac{{\nc}_{\bk-\bp}}{a_\bp} (\g{\bp}-\gd{-\bp}). 
\ee
Here, $\Nc = \int\rmd^dr \,\nc(\br) = L^d \nc$ is the total number of particles in the condensate with average density $\nc$.  

The pair-bogolon current \eqref{currentg2} reads 
\be\label{g2k} 
\hat\bg^{[2]}_\bk=\frac{1}{2L^d}\sum_\bp \bp \frac{a_{\bk-\bp}}{a_\bp}(\g{\bp}-\gd{-\bp})(\g{\bk-\bp}+\gd{\bp-\bk}). 
\ee 
For the purpose of evaluating \eqref{rhonggdef} in the limit $\bk\to0$,  
one may take $a_{\bk-\bp}/a_\bp \to 1$ in this expression right away, easing notations. 
We must, however, keep track of $\bk$ in $\g{\bk-\bp}$ and $\gd{\bp-\bk}$ to ensure their commutativity with $\gd{-\bp}$ and $\g{\bp}$, respectively. 
By virtue of symmetry under
$\bp\leftrightarrow \bk-\bp$, the pair-bogolon current \eqref{g2k} then  
simplifies to 
\be\label{g2}
\hat\bg^{[2]}_\bk \stackrel{\bk\to 0}{=} \frac{1}{L^d}\sum_\bp \bp \gd{\bp}\g{\bp+\bk}. 
\ee 
In this pleasing form, the pair-bogolon current operator is analogous to the standard single-particle
current for  $\bk\to0$, and one enjoys \resub{the} standard Feynman rules for
diagrammatic expansions
\cite{Mahan2000,Bruus2004}. 

The bogolon Matsubara-Green
function $G$ and its anomalous counterpart $F$ are  
\be\begin{aligned}
G_{\bk\bk'}(i\omega_n) = &-\int_0^\beta\rmd \tau e^{i\omega_n\tau}
\xpct{\g{\bk}(\tau)\gd{\bk'}(0)},\\
F_{\bk\bk'}(i\omega_n) = &-\int_0^\beta\rmd \tau e^{i\omega_n\tau}
\xpct{\gd{-\bk}(\tau)\gd{\bk'}(0)}. 
\end{aligned}
\ee
Just as the external potential 
$\calV
=\begin{smallpmatrix} W & 
  Y\\ Y & W\end{smallpmatrix}
$, they can be grouped in a Nambu matrix  
$\calG= 
\begin{smallpmatrix}
G & F^*\\
F & G^*
\end{smallpmatrix}$. 
For the \resub{translation-invariant} Bogoliubov Hamiltonian \eqref{H0}, $\calG_0 = \begin{smallpmatrix} 
G_0 & 0 \\
 0& G_0^*\end{smallpmatrix}$ is diagonal in Nambu \resub{pseudo-spin space ($F_{0}=0$)} and momentum representation, 
$G_{0\bk\bk'}(i\omega_n) =
\delta_{\bk\bk'}(i\omega_n-\ep{\bk})^{-1}$. 
In the presence of scattering by the external potential \eqref{calVdef}, the Nambu-Green matrix expands into the Born series 
\be\label{Born} 
\calG= \calG_0 +\calG_0\calV\calG_0+\calG_0\calV\calG_0\calV\calG_0
+\dots
\ee 
Since only elastic
scattering by a static potential is considered, all Green functions
in this expansion are taken at the same frequency.

When the current correlator \eqref{rhonggdef} is evaluated over a thermal state of the
quadratic Hamiltonian $\hat H=\hat H_0+\hat\calV$, Wick's theorem
guarantees that there are no 
mixed terms between the single- and pair-bogolon contributions: 
\be
\rhon= \rhon^{[1]} + \rhon^{[2]}. 
\ee
These contributions yield rather different responses and are calculated separately in the two
following sections. 

\section{Single-bogolon contribution} 
\label{g1.sec}

In a homogeneous system, condensation occurs in the 
$\bk=0$ mode, ${\nc}_\bk=\nc\delta_{\bk0}$. 
Then, the single-bogolon current \eqref{g1k} 
is purely longitudinal, $\hat\bg^{[1]}_\bk \propto
\bk$, and thus contributes nothing to the transverse current that defines the normal component \eqref{rhonggdef}.  
In presence of an external
potential, however, the condensate is deformed, and 
\eqref{g1k} acquires a transverse component.  
Its contribution to \eqref{rhonggdef} evaluates to 
\be\label{rhon1}
\rhon^{[1]} = \frac{1}{2\nc}\sum_{\bp,\bq} 
\frac{p_z q_z}{a_\bp a_\bq}
\avg{
\left.
{\nc}_{\bk-\bp}{\nc}_{\bq-\bk}
\right|_{\bk_\perp\to0} 
[F_{\bp\bq}(0) - G_{\bp\bq}(0)]
}
.
\ee 

\resub{This formula admits a straightforward perturbative expansion in powers of $V$ by expanding both the condensate densities
and Matsubara-Green's functions and collecting terms of the required order.} 
The condensate density expands in powers of the external potential 
as 
\cite{Sanchez-Palencia2006,Gaul2011a,Lugan2011,Mueller2012} 
\be\label{ncsmoothing}
{\nc}_\bk = \nc (\delta_{\bk0} - 2\wtV_\bk +\dots).
\ee 
The small parameters here are the effective potential Fourier components 
\be\label{wtV} 
\wtV_\bk = \frac{V_\bk}{\epn{\bk} + 2\mu}= \frac{V_\bk/\mu}{k^2\xi^2 + 2}
\ee
that encode the deformation of the condensate.
The spatial condensate deformation is considerably smoothed for random potentials 
with high spatial frequency $k \gg \xi^{-1}$
compared to the healing length $\xi =1/\sqrt{2m\mu}$ (remember our choice of units with $\hbar=1$), but 
follows the potential faithfully in the opposite, Thomas-Fermi limit
$k \ll \xi^{-1}$, where $\wtV_\bk \approx V_\bk/2\mu$.


To the lowest 
nontrivial order $V^2$ in the external potential, \eqref{rhon1} agrees with a well-known result. 
Indeed, the zeroth order of \eqref{ncsmoothing}, the homogeneous condensate, 
does not contribute because of the transversality constraint 
\resub{$\sum_\bp  p_z\delta_{\bk_\perp \bp} = 0$}. 
Therefore, the background densities are
at least first order 
in $V$. 
To the overall order $V^2$, only the free propagators $G_{0\bp\bq}(0) = -\delta_{\bp\bq}\ep{\bp}^{-1}$ and $F_{0\bp\bq}= 0$
need to be considered, and  
\eqref{rhon1} boils down to
\be\label{rhon1V2}
\rhon^{[1]} = 2\nc \sum_\bp \frac{p_z^2}{\epn{\bp}} \avg{|\wtV_\bp|^2}.  
\ee
When the \resub{effective} potential power spectrum $\avg{|\wtV_\bp|^2}$ is 
isotropic, this expression simplifies further, in $d$ dimensions 
\cite{Huang1992,Giorgini1994,Lopatin2002}
\be\label{rhon1V2iso}
\rhon^{[1]} =\frac{4\rhoc}{d}  \sum_\bp \avg{|\wtV_\bp|^2}, 
\ee 
with $\rhoc=m\nc$ the condensate mass density. 
The number $\sum_\bp \avg{|\wtV_\bp|^2} = V_2$ is 
the amount by which the external
potential to second order depletes the $\bk=0$ component
\cite{Lopatin2002,Mueller2012,Astrakharchik2013}, namely the coherent component of the condensate
\cite{Mueller2015}.  
\resub{It is, indeed, rather reasonable that the superfluid depletion links to the depletion of the coherent component (density of condensed particles with $k=0$), which must be distinguished from the overall condensate depletion, i.e., the density $\delta n = n-\nc$ of non-condensed particles irrespective of their momentum, which is finite even in a homogeneous system at zero temperature \cite{Penrose1956}. }

\begin{figure} 
\centering
\includegraphics[width=\linewidth]{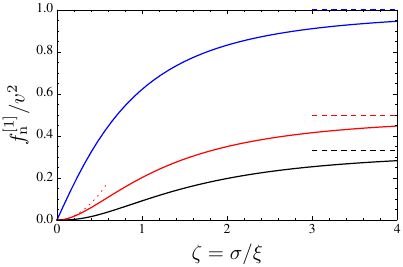}
\caption{Single bogolon contribution to the normal fraction, eq.~\eqref{fn1.eq}, to first order in the disorder potential variance $v^2=V^2/\mu^2$, as function of reduced potential correlation length $\zeta = \sigma/\xi$ in dimensions $d=1,2,3$ (top to bottom). Dashed: Thomas-Fermi limit $\fnTF{1}=v^2/d$.}
\label{fn1.fig}
\end{figure}

We briefly discuss some properties of \eqref{rhon1V2iso}, introducing notations that will be useful 
for similar analyses in Sec.~\ref{g2.sec} below.  
The external potential appears in \eqref{rhon1V2} and 
\eqref{rhon1V2iso} via \eqref{wtV} with its power spectrum  
\be \label{Cpsigma}
\avg{|V_\bp|^2} = V^2 (\sigma/L)^d C_d(\sigma \bp),
\ee 
\resub{the Fourier transform of the spatial pair correlator $C(\br-\br') =  \avg{V(\br)V(\br')}/V^2$, with}
prefactors chosen to ensure the normalization 
$(\sigma/L)^d\sum_\bp C_d(\sigma\bp)= \resub{C(0) ={} } 1$. 
For a given
reduced potential strength $v^2=V^2/\mu^2$, the single-bogolon contribution to the normal
fraction becomes   
\be\label{fn1.eq}
\fn^{[1]} = \frac{\rhon^{[1]}}{\rho} 
= \frac{4 v^2 }{d} \int\frac{\rmd^dx}{(2\pi)^d} \frac{\zeta^dC_d(\zeta\bx)}{(2+x^2)^2}, 
\ee
the integral being carried out over the reduced momentum $\bx = \xi\bp$. 
To leading order, 
$\rhoc\approx\rho$ has dropped out. 
The result   
\eqref{fn1.eq} is only a function of $\zeta = \sigma/\xi$, i.e., the correlation length of the
potential in units of the healing length. It is plotted in
Fig.~\ref{fn1.fig} for dimensions $d=1,2,3$,    
and a generic, isotropic Gaussian correlation  
$C_d(\bx) = (2\pi)^{d/2}\exp\{-\bx^2/2\}$. 
The general tendency is towards a larger superfluid depletion in lower
dimensions and for longer correlations, where the external potential is less screened by the condensate.

For a $\delta$-correlated potential, i.e., in the white-noise limit $\zeta=\sigma/\xi\to0$ and $v = V/\mu \to\infty$ with
$v^2\zeta^d=:v_\delta^2$ held constant, the normal fraction becomes  
$\fn^{[1]}= f_d v_\delta^2$ with $f_d= C_d(0)(2\pi)^{-d} \int \rmd^d x [2+x^2]^{-2}$ a
non-universal constant that depends on the long-range properties of
the potential correlator.   

In the opposite limit of a
very smooth potential, $\zeta\to\infty$, the potential correlator \eqref{Cpsigma}
tends to a delta distribution such that 
$\avg{|\wtV_\bp|^2} \to v^2\delta_{\bp0}/4$. 
Then, the normal fraction 
converges to the universal, Thomas-Fermi limit $\fnTF{1} = v^2/d$, independent of potential correlation details%
\resub{, and governed by the extreme phonon limit $\ep{\bp}\approx cp \to 0$ of the Bogoliubov dispersion, with $c=\sqrt{\mu/m}$ the sound velocity.} 

The single-bogolon contribution \eqref{rhon1} to the normal density 
is independent of temperature.  
Thus, Landau's finite-temperature relation \eqref{rhonLandau} is not
affected, and indeed not even reproduced for the clean system. 
In order to derive 
the finite-temperature superfluid depletion \eqref{rhonLandau}, one has two
choices: either introduce interactions among the quasiparticles \cite{Giorgini1994,Mora2003} or evaluate the
pair-bogolon current response in the transverse channel with the
quadratic Hamiltonian. 
The advantage of the latter choice is that the thermal average can be performed exactly using Wick's theorem, and this is the method we have chosen here in order to evaluate disorder corrections to eq.~\eqref{rhonLandau}. 
The drawback is that one can no longer rely on the
continuity equation and sum rules in the longitudinal channel, which are recalled in App.~\ref{longitudinal.sec}; similar
restrictions applied in Refs.~\cite{Huang1992,Giorgini1994}.

\section{Pair-bogolon contribution}
\label{g2.sec}

\begin{figure}
\centering
(a) \includegraphics[width=0.425\linewidth]{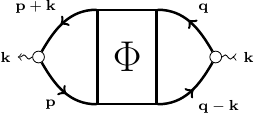}
(b) \includegraphics[width=0.425\linewidth]{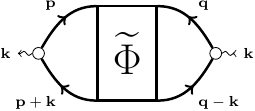}
\caption{
Pair contributions to the current-current correlations: 
(a) normal contribution, Eq.~\eqref{Phikernel}. 
(b) anomalous contribution, Eq.~\eqref{tildePhikernel}. 
}
\label{Phidiagfig}
\end{figure}

The pair-bogolon current contribution  
to the normal density \eqref{rhonggdef}  is the
disorder-averaged, zero-transverse momentum and zero-frequency limit, 
\be 
\rhon^{[2]}= \avg{\Upsilon_{zz}(\bk_\perp\to0,ik_n\to 0)}, 
\ee 
of the two-quasiparticle Green function 
\be\label{Upsilonzztau} 
\Upsilon_{zz}(\bk,\tau) = \frac{1}{L^d}\sum_{\bp,\bq} p_z q_z
\xpct{\mathrm{T}\gd{\bp}(\tau)\g{\bp+\bk}(\tau)\gd{\bq}(0)\g{\bq-\bk}(0)}.
\ee
Here, we have used the simplified current \eqref{g2}, $\mathrm{T}$ is the
(imaginary-)time ordering operator, and the average is taken over the
thermal equilibrium state 
$Z^{-1}\exp\{-\beta (\hat H_0+\hat\calV)\}$ of our quadratic Bogoliubov
Hamiltonian. 
The formalism here is very similar to the linear response of BCS
theory 
\cite{Mahan2000,Bruus2004}, 
and the results have many features in common. 

By virtue of Wick's theorem, this two-particle Green function 
is a sum of products of single-particle Green functions, 
\begin{multline}
  \xpct{\mathrm{T}\gd{\bp}(\tau)\g{\bp+\bk}(\tau)\gd{\bq}(0)\g{\bq-\bk}(0)} = \\
 G_{\bp+\bk,\bq}(\tau)G^*_{\bp,\bq-\bk}(\tau) 
+ F_{-\bp,\bq}(\tau)F^*_{-\bp-\bk,\bq-\bk}(\tau).  
\end{multline}
After transformation to Matsubara frequency
$ik_n\to0$, the static  
two-quasiparticle correlator, 
\be\label{Upszz}
\Upsilon_{zz}(\bk,0) =  \frac{1}{L^d}  \sum_{\bp,\bq} p_z
q_z [\Phi_{\bp\bq}(\bk,0) + \widetilde\Phi_{\bp\bq}(\bk,0)], 
\ee
features a normal and an anomalous contribution, 
\begin{eqnarray}
 \Phi_{\bp\bq}(\bk,0)  =& \frac{1}{\beta}\sum_{ip_n}
 G_{\bp+\bk,\bq}(ip_n)G_{\bq-\bk,\bp}(ip_n), \label{Phikernel} \\
 \widetilde\Phi_{\bp\bq}(\bk,0)  =& \frac{1}{\beta}\sum_{ip_n}
 F_{-\bp,\bq}(ip_n)F^*_{-\bq-\bk,\bp+\bk}(ip_n),
\label{tildePhikernel} \end{eqnarray} 
depicted in Fig.~\ref{Phidiagfig}(a) and (b), respectively. 
(We retain solely the connected contributions shown in
Fig.~\ref{Phidiagfig} because the disconnected contribution proportional to $G_{\bp+\bk,\bp}(0^-)G_{\bq-\bk,\bq}(0^-)$ vanishes.)

\begin{figure}
\centering
(a) \includegraphics[width=0.425\linewidth]{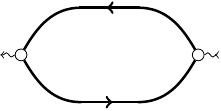}
(b) \includegraphics[width=0.425\linewidth]{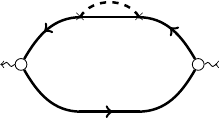}\\
(c) \includegraphics[width=0.425\linewidth]{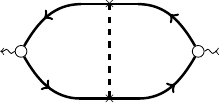}
(d) \includegraphics[width=0.425\linewidth]{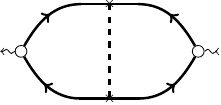}
\caption{(a) The current-current pair correlator of the clean system, Eq.~\eqref{Phiclean}, yields Landau's formula, Eq.~\eqref{rhonLandau}.
Two types of impurity corrections are found \resub{at order $\calV^2$}:  
Type I, shown in (b): self-energy correction to the free normal propagator. 
Up to order $\calV^2$, this does not affect the anomalous kernel, since $F_0$ on the other side vanishes. 
Type II:   
vertex corrections connect normal (c) and anomalous (d) propagators on opposite sides.  
}
\label{PhidiagV2.fig}
\end{figure}

\subsection{Clean system: Landau's formula recovered}

To check whether we are on the right track, we evaluate these expressions for the clean system where
$\calV=0$. There,  the anomalous Green's function vanishes, $F_0=0$, and so does the anomalous kernel, $\widetilde\Phi_0=0$.  The clean normal 
kernel \eqref{Phikernel}, shown in Fig.~\ref{PhidiagV2.fig}(a), evaluates at zero momentum to 
\be\label{Phiclean}
 \Phi_{0\bp\bq}(0,0) =\delta_{\bp\bq} \frac{1}{\beta} \sum_{ip_n}
 \frac{1}{(ip_n-\ep{\bp})^2} = -\delta_{\bp\bq}\partial_{\ep{\bp}}
 \nu_{\ep{\bp}}. 
\ee
Using these results in \eqref{Upszz} produces 
\be\label{rhon20}
\rhon = \rho^{[2]}_{\text{n}0} =  \frac{1}{L^d}\sum_\bp p_z^2 (-\partial_{\ep{\bp}}\nu_{\ep{\bp}}) 
\ee
which turns into Landau's formula \eqref{rhonLandau} in the thermodynamic limit. The resulting normal fraction is 
\be\label{fn0}
f_{\text{n}0} (T) = \frac{a_d}{n\xi^d}  \int_0^\infty \rmd x x^{d+1}
(-\partial_{\ep{x}}\nu_{\ep{x}})  
\ee
with $a_d = 4/[d (4\pi)^{d/2}\Gamma(d/2)]$. 
The integal runs over $x = p\xi$, i.e., momentum in units of $1/\xi$ and thus excitation energies in units of $\mu$, $\ep{x}=|x|(2+x^2)^{1/2}$. 
The dependence on temperature $\kB T/\mu =
1/\beta\mu$ arises from the bosonic thermal distribution $\nu_{\ep{}} =
[\exp\{\beta \mu\ep{}\}-1]^{-1}$. 
The quadratic Bogoliubov approximation holds as long
as this normal fraction is small. Since the mean-field, symmetry breaking description of the condensate requires $1/n\xi^d\ll1$, this can cover temperatures $\kB T > \mu$, but of course smaller than the BEC
transition temperature.  

In the low-temperature
regime $\kB T \ll \mu$, where the dispersion contributes with its phonon asymptotics $\ep{x}\approx
\sqrt{2}|x|$, \eqref{fn0} can be approximated by 
\cite{Pitaevskii2003} 
\be\label{lowTfn0} 
f_{\text{n}0}(T) \approx 
\frac{f_d}{n\xi^d}
\left(\frac{\kB T}{\mu}\right)^{d+1}
\qquad (\kB T \ll \mu), 
\ee
with 
$f_d 
= 2\Gamma(d+2) \zeta (d+1)/[d(8\pi)^{d/2}\Gamma(d/2)]$.

\subsection{Inhomogeneous system: corrections $O(V^2)$}

Using the Born series \eqref{Born}, we can identify impurity-scattering corrections 
to the free Green's
functions. 
There are two types of
corrections of order $\calV^2$, shown in Fig.~\ref{PhidiagV2.fig}:
(I) self-energy insertions to one of the propagators [Fig.~\ref{PhidiagV2.fig}(b)], and (II) vertex corrections connecting 
opposite propagators [Fig.~\ref{PhidiagV2.fig}(c) and (d)]. 
Up to second order in $\calV$, self-energy insertions for
the anomalous propagators need not be considered since the other,
free anomalous propagator $F_0$ vanishes anyway.  
Whereas some self-energy corrections have been considered in
Ref.~\cite{Lopatin2002}, to our knowledge the 
vertex corrections have not been identified and computed so far. 
\resub{These two types of impurity-scattering diagrams must always be considered together in order to satisfy basic conservation laws, as expressed by so-called Ward identities \cite{Mahan2000,Bruus2004}, linking the (irreducible single-particle) self-energy corrections to the (irreducible two-particle) vertex corrections. 
The two diagrams shown in Fig.~\ref{PhidiagV2.fig}(b) and (c) are the most elementary example of such a link, featuring the only corrections of order $\calV^2$ (the so-called first-order Born approximation) that can be applied to the free single-quasiparticle propagator on the one hand and the two-quasiparticle correlator on the other. 
For higher-order diagrams or self-consistent approximation schemes, Ward identities are a crucial ingredient \cite{Vollhardt1980}, and clearly applicable, since our Hamiltonian has the generic structure $\hat H= \hat H_0 +\hat\calV$ of single-particle scattering by static impurities.  
}

\resub{In the present work, we limit ourselves to a perturbative calculation in the bare external potential $V$ to the lowest order, so there is no need to consider higher-order diagrams.} To the quadratic order of interest, we first select all contributions
$\calV,\calV^2$ \resub{arising from the Born series \eqref{Born}} and then expand these in powers of the bare
potential $V$ \resub{(see the appendix \ref{pertV.sec} for details)}.
Collecting all terms of order $V^2$ finally yields the
pair-bogolon contribution to the normal density
\be\label{rhon2}
\rhon^{[2]} = \frac{1}{L^d}\sum_{\bp,\bp'} \Pi_{\bp\bp'}\avg{|\wtV_{\bp-\bp'}|^2}
\ee
where
\be
\begin{split}\label{Rpq}
\Pi_{\bp\bp'} 
  = &  \frac{1}{2}
| y^{(1)}_{\bp\bp'}|^2 (p_z\partial_{\eps}-p'_z\partial_{\eps'})^2
\frac{1+\nu_\eps +\nu_{\eps'}}{\eps+ \eps'}  \\
 & - \frac{1}{2}|w^{(1)}_{\bp\bp'}|^2 (p_z\partial_{\eps}+p'_z\partial_{\eps'})^2
\frac{\nu_{\eps}-\nu_{\eps'}}{\eps- \eps'} \\
 & + w^{(2)}_{\bp\bp'}p_z^2 (-\partial_\eps^2\nu_\eps)
\end{split}
\ee
with the short-hand notation $\eps=\ep{\bp}$, $\eps'=\ep{\bp'}$. 
The envelope matrix elements $y^{(1)}_{\bp\bp'}$, $w^{(1)}_{\bp\bp'}$, and $w^{(2)}_{\bp\bp'}$ are 
\resub{detailed in appendix \ref{pertV.sec} in eqs. \eqref{w1}, \eqref{y1}, and \eqref{w2}, respectively. }

Expression \eqref{Rpq} appears intimidating, but is
relatively simple to explain. 
The contribution of the last term proportional to $w^{(2)}$ is
actually the easiest to understand:
Due to the nonlinear condensate deformation, the bogolon self-energy
contains a non-resonant term $W^{(2)}$ of order $V^2$, without intermediate propagator 
(\resub{for a diagrammatic representation, see eqs.~(33) and (69) in} \cite{Gaul2011a}),    
describing a purely real dispersion shift 
$\eps\to\eps + W^{(2)}$. 
Thus, the distribution function is Taylor-expanded $\nu_\eps \to \nu_\eps+ W^{(2)} \partial_\eps\nu_\eps $, which explains the corresponding disorder correction to the Landau formula \eqref{rhon20}.   

The first and second term in Eq.~\eqref{Rpq} group all the corrections that involve   
intermediate normal and anomalous propagators, respectively, in the
diagrammatic expansion. A first
contribution to such terms comes from the self-energy insertions of type
I shown in Fig.~\ref{PhidiagV2.fig}(b). 
A second contribution comes from the vertex
corrections of type II, shown in Fig.~\ref{PhidiagV2.fig}(c) and (d). 
The bogolon occupation
numbers enter in combinations that are characteristic for normal and
anomalous channels, just as expected from the formal similarity with BCS theory (see, e.g., Eq.~(18.66) in
Ref.~\cite{Bruus2004});
similar expressions appear in the calculation of the condensate depletion at finite temperature \cite{Gaul2014}. 
Only the correct combination of self-energy and vertex insertions, all
to the same order in $V$, results in the remarkably well-behaved Eq.~\eqref{Rpq}.

\begin{figure}
\centering
\includegraphics[width=\linewidth]{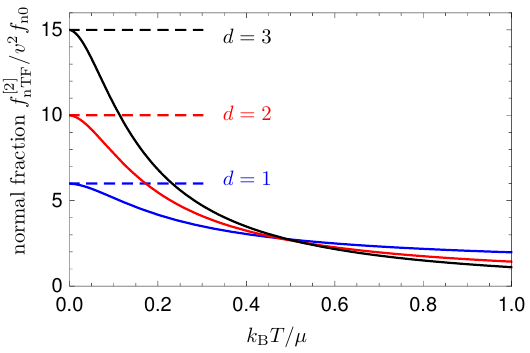}
\caption{Finite-temperature normal fraction in the Thomas-Fermi regime
  of smooth potentials, Eq.~\eqref{fnTF2}, per
  potential variance $v^2=V^2/\mu^2$ and   
relative to the clean fraction, Eq.~\eqref{fn0}, as function of reduced temperature $\kB
  T/\mu$. Towards zero temperature, their ratio becomes constant, Eq.~\eqref{fnTFlowT.eq}, shown as dashed lines.   
 }
\label{fnTF.fig}
\end{figure}

In order to gain a better understanding of the temperature-dependent disorder correction, we study it in an analytically tractable limit that is also relevant for cold gases in optical potentials, namely the Thomas-Fermi regime $\zeta=\sigma/\xi\to\infty$. 
There, the potential correlator $\avg{|\wtV_\bq|^2}\to v^2\delta_{\bq0}/4$ contracts one of the two sums in Eq.~\eqref{rhon2} and yields the much
simpler result 
\be
\rhonTF^{[2]} = \frac{v^2}{4 L^d}\sum_\bp \Pi_{\bp\bp}. 
\ee
Thus, the term proportional to $|y^{(1)}_{\bp\bp}|^2$ in Eq.~\eqref{Rpq} (supporting in the general case a small 
temperature-independent correction, see App.~\ref{TindepPairContrib.sec}) vanishes by symmetry. 
The remaining terms involving $w^{(2)}_{\bp\bp}$ and $|w^{(1)}_{\bp\bp}|^2$ simplify somewhat, and the
induced normal fraction writes
\be\label{fnTF2}
\fnTF{2} (T) = \frac{a_d v^2}{2 n\xi^d} 
\int_0^\infty \rmd x\frac{x^{d+5}}{\ep{x}}
\left[-\partial_{\ep{x}}^2\nu_{\ep{x}} 
 - \frac{4}{\ep{x}}\partial_{\ep{x}}^3\nu_{\ep{x}} \right]. 
\ee

 
The second term in the brackets, which 
stems from the $w^{(1)}$-contribution, is positive.  
In the low-temperature limit $\kB T\ll\mu$, it contributes with the same temperature and phase-space dependence
$\propto  (n\xi^d)^{-1} (\kB T/\mu)^{d+1}$ 
\resub{as} 
the clean
depletion $f_{\text{n}0}(T)$, Eq.~\eqref{lowTfn0}. The disorder-affected, temperature-dependent
normal fraction in the Thomas-Fermi regime can then be written 
\be\label{fnTFlowT.eq}
\fnTFempty (T) \approx  f_{\text{n}0}(T) \left[1+\frac{\Gamma(d+4)}{2\Gamma(d+2)}
  \frac{V^2}{\mu^2} \right], 
\ee
up to higher orders in $V/\mu$ and $\kB T/\mu$ within the brackets.

The first term in the brackets of Eq.~\eqref{fnTF2}, which stems from the self-energy insertion $\propto w^{(2)}$, is a strictly negative contribution, 
but smaller by a factor $(\kB T/\mu)^2$ in the low-temperature limit. 
As apparent in Fig.~\ref{fnTF.fig}, where the disorder-induced, $T$-dependent normal fraction is plotted relative to the clean normal fraction, the $w^{(2)}$-term then reduces the relative correction $\fnTF{2}/f_{\text{n}0}$ quite drastically as the temperature rises.

\section{Summary and Outlook} 

Based on inhomogeneous Bogoliubov theory, applicable to weakly interacting,
(quasi-)condensed Bose gases, we have calculated the normal fraction of a superfluid of arbitrary dimensionality in a static, disordered potential by a diagrammatic evaluation of the transverse current correlator. 
This approach succeeds 
\resub{in keeping} track of finite spatial correlations in the external potential and 
\resub{in accounting} for all effects of disorder to the desired order in potential strength.  
Closed analytical
expressions are obtained for the normal density to second order in the
external potential strength. We have found simple analytical results in the
Thomas-Fermi regime of smooth potentials, whose correlation length is
larger than the condensate healing length. 

As a perturbative approach, these results cannot be naively extrapolated to stronger disorder and thus cannot be expected to  predict accurately the critical disorder strength at which the superfluid undergoes the phase transition to the normal (insulating Bose glass) phase, although promising attempts in this direction are on record \cite{Wang2016}.   
It would be interesting, however, to investigate whether certain, maximally crossed classes of impurity diagrams, which are known to be mainly responsible for Anderson localization of single particles \cite{Vollhardt1980}, play a dominant role at suppressing superfluidity by inhibiting the transport of elementary excitations in strongly disordered Bose-Einstein condensates \cite{Zuninga2013}. This question must be left for future research.

\begin{acknowledgments}
\resub{This work is dedicated to the memory of Dominique Delande.} The author gratefully acknowledges invited professorships at Universit\'e Pierre et Marie Curie (now Faculté des Sciences et Ingénierie de Sorbonne Université, Paris, France) and Universit\'e C\^ote d'Azur and CNRS, Nice, France, where this work was started.
\end{acknowledgments}

\appendix 

\section{Inhomogeneous Bogoliubov theory}
\label{InhBogoTheory.sec}

We recall, and summarise, the derivation of the inhomogeneous Bogoliubov Hamiltonian \eqref{InhBogH}, as well as the perturbative expansion of the bogolon scattering matrix $\calV$ in powers of the bare external potential $V$.  

\subsection{Quadratic Hamiltonian for the excitations of an inhomogeneous condensate}

A gas of massive Bosons with short-range interaction subject to a one-body potential $V(\br)$ is described by 
the Hamiltonian (in units such that $\hbar=1$) \cite{Dalfovo1999,Pethick2002,Pitaevskii2003}
\be\label{eqManyParticleHamiltonian}
 \hat H = \intddr\,
\hat \psi(\br)^\dagger \left[-\frac{\nabla^2}{2m}  + V(\br) - \mu + \frac{g}{2}\hat \psi(\br)^\dagger
 \hat\psi(\br)
 \right] \hat \psi(\br),
\ee
$\hat\psi(\br)$ and $\hat\psi^\dagger(\br)$ being the bosonic annihilation and creation operator, respectively. 
Below a critical temperature, the Bose gas forms a BEC \cite{Einstein1925}, with a macroscopically large fraction of particles populating the ground state of the single-particle density matrix \cite{Penrose1956}.
Following Bogoliubov \cite{Bogoliubov1947}, we split the field operator into the macroscopically occupied mean-field condensate and quantized fluctuations:
\begin{equation}\label{eqBogoliubovPrescription}
\hat \psi(\br) = \Phi(\br) + \delta\hat\psi(\br) .
\end{equation}
This approach violates strict number-conservation of particles and thus can only be approximately valid as long as the number of excited particles is small compared to the number of condensed particles. 
For dilute condensed atomic gases at low temperatures, Bogoliubov theory proves to be versatile and adequate in many circumstances. 
In particular, within confining potentials quasi-condensates can exist in dimensions smaller than $d=3$, provided the phase coherence length exceeds the system size $L$ \cite{Mora2003}. 

The condensate amplitude $\Phi(\br)$ minimizes the energy functional 
\begin{widetext} 
\begin{align}\label{eqGPEnergyFunctionalPsi}
E [\Phi(\br)]= \intddr \biggl\lbrace
\frac{1}{2m} |\vc{\nabla} \Phi(\br)|^2 + \big[ V(\br)- \mu \big] |\Phi(\br)|^2
+\frac{g}{2}|\Phi(\br)|^4	\biggr\rbrace .
\end{align}
\end{widetext} 
The stationarity condition $\delta E/\delta\Phi^*=0$ yields the time-independent Gross-Pitaevskii (GP) equation
\cite{Pitaevskii1961,Gross1963} 
\begin{equation} \label{eqStationaryGPE} 
-\frac{1}{2m} {\nabla^2\Phi(\br)} + \left(g |\Phi(\br)|^2 - \mu\right) {\Phi(\br)} =
- V(\br) {\Phi(\br)}  .
\end{equation}
Within this mean-field description, the condensate spontaneously breaks the U(1) gauge invariance of the Hamiltonian \eqref{eqManyParticleHamiltonian}. In a time-independent potential $V(\br)$, the condensate's kinetic energy is minimized by choosing a fixed global phase, and without loss of generality one can take $\Phi(\br) \in \mathbb{R}$.

Inserting the Bogoliubov ansatz \eqref{eqBogoliubovPrescription} into the Hamiltonian \eqref{eqManyParticleHamiltonian}, we expand it in powers of $\delta \hat\psi$ and $\delta \hat\psi^\dagger$. 
To zeroth order, of course, we find the GP energy functional \eqref{eqGPEnergyFunctionalPsi}.
The linear term vanishes, by construction, because $\Phi(\br)$ minimizes this energy. 
The first relevant contribution is then a Hamiltonian that is quadratic in the excitations. 
Terms of third and forth order are beyond quadratic Bogoliubov theory; they describe interactions between the excitations that become only relevant for larger densities or higher temperatures \cite{Katz2002}. 

As explained in \cite{Gaul2011a}, is is advisable to use the canonically conjugate density-phase variables, obeying 
$\bigl[\delta \hat n(\br), \delta \hat \varphi(\br') \bigr] = i \delta(\br-\br')$, 
thus ensuring the all-important (bi-)orthogonality of the excitations and their orthogonality to the inhomogeneous background $\Phi(\br) =\sqrt{\nc(\br)}$. 
(An alternative, but equivalent derivation bypassing the density-phase variables can be found in \cite{Mueller2012}.)
Inserting 
\be
\begin{split} 
\hat\psi(\br) & = \exp\{i\delta\hat\varphi(\br) \} \sqrt{\nc(\br)+\delta \hat n(\br)}\\
&= \Phi(\br)+ \delta\hat n(\br)/2\Phi(\br) +i\Phi (\br)\delta\hat \varphi(\br)+\dots 
\end{split} 
\ee 
and expanding \eqref{eqManyParticleHamiltonian} to second order, the quadratic fluctuation Hamiltonian is found to read
\cite{Gaul2008}
\begin{widetext} 
\begin{align}
\hat H_2  
= \int {\rm d}^d r \biggl\lbrace 
\frac{1}{2m} \biggl[  &\left(\nabla \frac{\delta \hat n(\br)}{2\Phi(\br)} \right)^2
+ \frac{\left[\nabla^2 \Phi(\br)\right]}{4 \Phi(\br)^{3}} \delta \hat n(\br)^2 
 + \Phi(\br)^2 \left(\nabla \delta \hat \varphi(\br)\right)^2\biggr] + \frac{g}{2} \,\delta \hat n(\br)^2 \biggr\rbrace .\label{eqBogoliubovHamiltonian}
\end{align}
\end{widetext}
To this order, there are no mixed terms of $\delta \hat n(\br)$ and $\delta \hat \varphi(\br)$.
The potential $V(\br)$ now enters only indirectly, and nonlinearly, via the condensate amplitude
$\Phi(\br)$. Since we aim at the weak-disorder regime $V\ll\mu = gn \approx g\nc$, the condensate can be safely assumed to be non-fragmented and vortex-free, with $\Phi(\br)>0$ everywhere inside the bulk.

Going over to the Fourier representation, the quadratic Hamiltonian remains quadratic, of course, with the general structure 
\begin{align}\label{eqEnergyFunctionalStructure}
\hat H_2 
&= \frac{1}{2}\sum_{\bk,\bk'}
\biggl\lbrace
 \delta \hat \varphi_\bk^\dagger S_{\bk\bk'} \delta \hat\varphi_{\bk'}
+  \delta \hat n_\bk^\dagger  R_{\bk \bk'} {\delta \hat n_{\bk'}}
 \biggr\rbrace.
\end{align}
The phase fluctations are coupled by  
\begin{align}\label{eqScattSgeneral.fields}
 S_{\bk\bk'} 
= \frac{1}{mL^d} \bk\cdot\bk' \sum_{\bp}  \Phi_{\bk-\bp}\Phi_{\bp-\bk'}, 
\end{align}
namely the convolution of condensate Fourier components $\Phi_{\bk}$ multiplied by $\bk \cdot\bk'$ stemming from the gradients.
The density fluctuation $\delta\hat n$ appears in \eqref{eqBogoliubovHamiltonian} in several, complicated looking terms. But the corresponding
coupling matrix $R_{\bk\bk'}$ can be
brought into a manageable form by   
using the Fourier components $\check\Phi_{\bk}$   
of the inverse amplitude $\check\Phi(\br) = \nc/\Phi(\br)$: 
\begin{widetext} 
\begin{equation}\label{eqScattRgeneral}
 R_{\bk \bk'} =  \frac{g\xi^2}{2\nc L^d} \sum_{\bp} \left[ p^2 + 2(\bk'-\bp)\cdot(\bk-\bp) + \frac{1}{2}(\bk'-\bp)^2+\frac{1}{2}(\bk-\bp)^2\right]  
 \check\Phi_{\bk-\bp}\check\Phi_{\bp-\bk'} + g \delta_{\bk\bk'}.
\end{equation}
\end{widetext}

Although the quadratic Hamiltonian \eqref{eqEnergyFunctionalStructure} appears diagonal, density and phase fluctuations are still coupled 
by their commutator
\begin{equation}\label{commutation_deln_delphi_k.eq} 
[\delta\hat n_\bk,\delta\hat\varphi_{\bk'}^\dagger]=i\delta_{\bk\bk'}.  
\end{equation}
This is resolved by a Bogoliubov transformation \cite{Bogoliubov1947}, combining  density and phase fluctuations judiciously into quasiparticle creation and annihilation operators $\gd{\bk}$  and $\g{\bk}$:
\begin{equation}\label{eqBgTrafo1}
\cvect{\g{\bk}}{\gd{-\bk}}
 = A_\bk 
\cvect{i\sqrt{\nc}\,\delta \hat \varphi_{\bk}}{\delta \hat
  n_\bk/2\sqrt{\nc}}, \quad A_\bk  =
\matr{a_\bk}{a_\bk^{-1}}{-a_\bk}{a_\bk^{-1}}. 
\end{equation}
A canonical transformation of this kind, with the free parameter $a_\bk$, guarantees that the quasiparticles obey bosonic commutation relations 
\begin{align}\label{eqCommutatorGamma}
 \bigl[\g{\bk},\gd{\bk'}\bigr] &= \delta_{\bk \bk'}, &
 \bigl[\g{\bk},\g{\bk'}\bigr] =  \bigl[\gd{\bk},\gd{\bk'}\bigr] &= 0.
\end{align}
We then separate the translation-invariant (homogeneous) contribution from the inhomogeneous 
contribution in the coupling matrices, $S_{\bk\bk'} = S_{\bk\bk'}^{(0)} +
S_{\bk\bk'}^{(V)}$ and $R_{\bk\bk'} = R_{\bk\bk'}^{(0)} +
R_{\bk\bk'}^{(V)}$. 
The `free' Hamiltonian $\hat H_2^{(0)}=:\hat H_0$ (we align the notations with Sec.~\ref{formalism.sec}) becomes diagonal,  
\begin{equation}\label{eqBgHamiltonianDiagonal}
\hat H_0 = \sum_{\bk} \, \ep{\bk} \gd{\bk}\g{\bk} ,
\end{equation}
by choosing 
\begin{equation}\label{eq_ak}
a_\bk = \left(\frac{\epn{\bk}}{\ep{\bk}}\right)^{1/2} = 
\left(\frac{k^2\xi^2}{2+k^2\xi^2}\right)^{1/4} .
\end{equation}
The free quasiparticles are thus found to have the famous Bogoliubov dispersion \cite{Bogoliubov1947}
\begin{equation}\label{eqBgDispersion}
\ep{\bk} = \sqrt{\epn{\bk}(2\mu+\epn{\bk})} = \mu \, k \xi \sqrt{2+k^2\xi^2}. 
\end{equation}
The same transformation defines the effective quasiparticle 
scattering vertex
\begin{equation}\label{eqDefWY}
\frac{(A_\bk^{-1})^\text{t}}{2}
\matr{S_{\bk\bk'}^{(V)}}{0}{0}{R_{\bk\bk'}^{(V)}} 
\frac{A_{\bk'}^{-1}}{2}
 = \matr{W_{\bk\bk'}}{Y_{\bk\bk'}}{Y_{\bk\bk'}}{W_{\bk\bk'}} 
=:\calV_{\bk\bk'}. 
\end{equation}
This finally achieves the derivation of the inhomogeneous Bogoliubov Hamiltonian:
\begin{equation}\label{eqInhomBgHamiltonian_gamma}
\hat H =  \hat H_0+\hat \calV  =  
\sum_{\bk} \, \ep{\bk} \gd{\bk}\g{\bk} +
\frac{1}{2} \sum_{\bk,\bk'} 
(\gd{\bk},\g{-\bk})  \calV_{\bk\bk'} \cvect{\g{\bk'}}{\gd{-\bk'}}.
\end{equation}
The only approximation, after the symmetry-breaking step, is the neglect of third- and forth-order terms in the fluctuations. 
Notably, by construction, the effective scattering potential $\hat\calV$ still encodes the complete information about the mean-field condensate deformation caused by the external potential \cite{Mueller2012}. 
Separating the `free' Hamiltonian $\hat H_0$ is structurally helpful in the sense that one can then apply standard recipes of diagrammatic quantum transport theory, such as a systematic expansion in $\hat\calV$. 
This does not mean, however, that applying these recipes is a trivial task. 
The scattering matrix $\calV_{\bk\bk'}$ as defined by \eqref{eqDefWY} is
a much richer object than a standard single-particle scattering amplitude  
$V_\bq$: First, it is
nonlinear in the bare external potential $V(\br)$ because the underlying
condensate -- solution of the GP equation \eqref{eqStationaryGPE} -- depends nonlinearly on the potential. 
Second, it has off-diagonal contributions $Y_{\bk\bk'}$ in the quasiparticle-pseudospin Nambu space because the underlying condensate 
mediates anomalous scattering between quasiparticle excitations.

\subsection{Perturbative expansion in $V$} 
\label{pertV.sec}

A well-defined pertubative expansion in powers of the bare potential $V\ll\mu$ up to a desired order $s$ ensues by the following steps: 
\begin{enumerate}
\item The imprint of the external potential on the condensate is perturbatively computed via the GP equation \eqref{eqStationaryGPE} \cite{Sanchez-Palencia2006} by expanding 
\begin{align}\label{eqSmoothingExpansionGPGs}
 \Phi_\bk =  \Phi^{(0)}_\bk + \Phi^{(1)}_\bk+ \Phi^{(2)}_\bk + \ldots + \Phi^{(s)}(\br)
\end{align}
around the homogeneous solution $\Phi^{(0)} =\sqrt{\nc}$, each contribution $\Phi^{(l)}(\br)$ being of order $l$ in 
the small parameter $V/\mu \ll 1$ \cite{Gaul2011a}. 

\item The Fourier amplitudes $\check\Phi_\bk$ of the inverse condensate amplitude $\check \Phi(\br) = \nc/\Phi(\br)$ are expanded to order $s$.  

\item 
The coupling matrixes $S_{\bk\bk'}$ and $R_{\bk\bk'}$ defined in \eqref{eqScattSgeneral.fields} and \eqref{eqScattRgeneral} are expanded to order $s$.

\item The Bogoliubov transformation \eqref{eqDefWY} is applied to obtain the quasiparticle scattering matrix 
\be 
\calV_{\bk\bk'} = \calV_{\bk\bk'}^{[1]} + \calV_{\bk\bk'}^{[2]} + \dots + \calV_{\bk\bk'}^{[s]}  
\ee
up to the desired order $s$ in $V/\mu$. 

\end{enumerate}
Following this straightforward procedure, one finds that the first-order scattering amplitudes 
\begin{align} 
W^{(1)}_{\bk \bk'} = w^{(1)}_{\bk \bk'} \wtV_{\bk-\bk'},
\qquad 
Y^{(1)}_{\bk \bk'} =  y^{(1)}_{\bk \bk'}  \wtV_{\bk-\bk'}
\end{align}  
are proportional to the reduced potential amplitudes $\wtV_\bp = V_\bp/\left(\epn{\bp}+2\mu\right)$, see eq. \eqref{wtV}, with envelopes 
\cite{Gaul2008,Gaul2011a,Mueller2012,Gaul2014}
\begin{align}
w^{(1)}_{\bk \bk'} &=& \frac{1}{2m}
\left[a_\bk a_{\bk'}( k^2 + k'^2 -\bk\cdot\bk') 
- a_\bk^{-1} a_{\bk'}^{-1}\, \bk\cdot\bk'   \right],   \label{w1}  \\ 
y^{(1)}_{\bk \bk'} &=& \frac{1}{2m}\left[ a_\bk a_{\bk'} ( k^2 + k'^2 -\bk\cdot\bk'
) + a_\bk^{-1} a_{\bk'}^{-1}\, \bk\cdot\bk'  \right].  \label{y1}
\end{align}
Second-order scattering amplitudes, as self-energy insertions for single-particle propagators, are only needed for 
$\bk=\bk'$: 
\be
W^{(2)}_{\bk \bk} = Y^{(2)}_{\bk \bk} 
 = \sum_{\bk'}  w^{(2)}_{\bk \bk'}  |\wtV_{\bk-\bk'}|^2
\ee 
with   
\be 
w^{(2)}_{\bk\bk'}  = \frac{1}{m} a_\bk^2 [k^2+(\bk-\bk')^2].
\label{w2}
\ee


\section{Continuity equation and longitudinal current response}
\label{longitudinal.sec}

Local number conservation, expressed by the continuity equation,
provides a sum rule for the longitudinal current response
function \cite{Baym1969,Pitaevskii2003,Ueda2010}. 
This sum rule can be used to derive 
a Josephson relation for disordered superfluids valid upon ensemble averaging \cite{Mueller2015}. 
Since the symmetry-breaking version of Bogoliubov theory violates
strict number conservation, it is appropriate to check how much of the
longitudinal sum rule survives. 

\subsection{Number-conserving theory} 
\label{numberconserving.sec} 

Given a Hamiltonian \resub{such as \eqref{eqManyParticleHamiltonian}}, of the form 
$\hat H[\hat\psi,\hat\psi^\dagger]=\hat K+ \hat U$  with kinetic energy 
\be
\hat K = \frac{1}{2m} \int \rmd^dr\, \nabla\hat \psi^\dagger(\br)\cdot\nabla\hat\psi{\br}
\ee 
and an interaction $\hat U=U[\hat\rho]$ that is a functional of the mass density $\hat\rho(\br)=m\hat\psi^\dagger(\br)\hat\psi(\br)$ 
only, 
the \resub{imaginary-time} equation of motion 
$\partial_\tau\hat\rho = [\hat H,\hat\rho] = [\hat K,\hat\rho]$ takes the form of the continuity equation 
\be\label{continuity} 
i\partial_\tau\hat\rho(\br,\tau)+\nabla\cdot\hat\bg(\br,\tau)=0, 
\ee 
with $\hat\bg(\br)=\frac{1}{2i}\left[\hat\psi^\dagger(\br)\nabla \hat\psi(\br) - \left(\nabla\hat\psi^\dagger(\br)\right)\hat\psi(\br) \right]$ the momentum (or mass current) density. 
In momentum representation 
\begin{align}
\hat\rho_\bk &= \frac{m}{L^d} \sum_\bp \hat a^\dagger_\bp \hat a_{\bp+\bk},  \\
\hat\bg_\bk &= \frac{1}{L^d} \sum_\bp \left(\bp+\frac{\bk}{2}\right)\hat a^\dagger_\bp \hat a_{\bp+\bk},
\end{align}
the continuity equation \eqref{continuity} becomes 
\be
\partial_\tau \hat\rho_\bk(\tau) + \bk \cdot\hat\bg_\bk(\tau) = 0,  
\ee
and thus links the density evolution to the longitudinal current,   
$\partial_\tau \hat\rho_\bk(\tau) = \bk \cdot\hat\bg_\bk(\tau) = - k \hat g_\bk^\parallel(\tau)$. 
Read from right to left, this relation permits us to evaluate the
zero-frequency longitudinal current response, because 
\be
\int_0^\beta\rmd\tau \xpct{\hat g_{\bk}^\parallel(\tau) \hat
  g_{-\bk}^\parallel(0) } = 
k^{-1} \xpct{[\hat\rho_\bk,\hat g_{-\bk}^\parallel]}  
\ee
then only involves the equal-time commutator 
\be 
[\hat\rho_\bk,\hat g_{-\bk}^\parallel] 
= \frac{k m}{ L^{2d}}
\sum_\bp \hat a^\dagger_\bp \hat a_\bp, 
= \frac{k}{ L^d}
\hat \rho_0, 
\ee
such that indeed 
\be\label{rho_sumrule}
L^d \int_0^\beta \rmd
\tau\xpct{\hat{g}^{\parallel}_{\bk}(\tau)\hat{g}_{-\bk}^\parallel(0)}
=  \xpct{\hat \rho_0} = \rho 
\ee 
yields the average mass density, 
independently of $\bk$. 

\subsection{Bogoliubov theory} 
\label{nonconserving.sec} 

Also in the Bogoliubov setting, the continuity equation can be used
to actually define the current linked to the density once the Hamiltonian is
known. Indeed, to lowest, quadratic order in
the fluctuations  $\dph$ and $\dn$, their \resub{Hamiltonian \eqref{eqBogoliubovHamiltonian}} reads  
\be
\hat H_2= \frac{1}{2m} \intddr \, \nc(\br)(\nabla\dph(\br))^2 +
U_2[\dn]. 
\ee
Thus, the equation of motion in form of the continuity equation 
\be\label{contBogoOne} 
im\partial_\tau\dn(\br,\tau) + \nabla \cdot \hat\bg^{[1]}(\br,\tau) =0
\ee
links the evolution of density fluctuations \resub{to the single-bogolon} current induced by the phase fluctuations,
as defined by Eq.~\eqref{currentg1}. 

One must be aware that standard Bogoliubov theory above a mean-field background violates strict number conservation, which limits
the applicability of the continuity equation for sum rules. Indeed, using the momentum version of the continuity relation  
\eqref{contBogoOne}
in the longitudinal current-current response, we find 
\be
\int_0^\beta\rmd\tau
\xpct{\hat g_{\bk}^{[1]\parallel}(\tau) \hat g_{-\bk}^{[1]\parallel}(0)} = 
\frac{m}{k L^{d/2}} \xpct{[\dn_\bk,\hat g_{-\bk}^{[1]\parallel}]}.  
\ee
The equal-time commutator is elementary to evaluate using the current \eqref{currentg1} in
the form 
\be
\hat\bg^{[1]}_\bk = \frac{i}{L^{d/2}}\sum_\bp \bp {\nc}_{\bk-\bp} \dph_\bp
\ee
since  $[\dn_\bk,\dph_{-\bp}]=i\delta_{\bk\bp}$, and we find as a result 
that the longitudinal current-current correlator yields
\be
L^d\int_0^\beta\rmd\tau
\xpct{\hat g_{\bk}^{[1]\parallel}(\tau) \hat g_{-\bk}^{[1]\parallel}(0)} = m \nc =\rhoc. 
\ee
\resub{Since the total density must be the sum $n = \nc +\delta n$ of the condensate density plus the non-condensed (or depleted) density, this result obviously misses the well-known condensate depletion (caused by interactions and finite temperatures as well as external potentials, and accessible to inhomogeneous quadratic Bogoliubov theory \cite{Mueller2012,Gaul2014}).     
Indeed,} 
the higher-order, pair-bogolon current, eq.~\eqref{g2k}, 
is linked to the dynamics of $\dn$ generated by a  
Hamiltonian that is cubic in the fluctuations, but which is only known in a homogeneous setting 
\cite{Mora2003}.
In the present, inhomogeneous setting, we are limited to the analytically known, quadratic
Hamiltonian \eqref{eqInhomBgHamiltonian_gamma}. 
As a consequence, we cannot rely on the continuity equation in order to 
evaluate expectation values of the longitudinal current
operator. Instead, the strategy of this work 
is to evaluate the transverse current
correlator \eqref{rhonggdef} directly using the quadratic
Hamiltonian \cite{Giorgini1994}. 

\section{Temperature-independent, pair-bogolon correction} 
\label{TindepPairContrib.sec}

At zero temperature where $\nu_{\eps}\to0$, a single term  
proportional to 
$| y^{(1)}_{\bp\bp'}|^2$ 
survives in the first line of \eqref{Rpq}, and the pair-bogolon contribution of order $v^2=V^2/\mu^2$ to 
the normal fraction becomes 
\be\label{fnzero2.eq}
\fn^{[2]}(0) = \frac{\rhon^{[2]}}{\rho} = \frac{v^2}{n\xi^d} \int
\frac{\rmd^d y}{(2\pi)^d}y_z^2 R_d(y)
\zeta^d C_d(\zeta \by)
\ee
with the 
isotropic kernel 
\be\label{Ndkernel} 
R_d(y) = \frac{2}{(2+y^2)^2} \int \frac{\rmd^d x}{(2\pi)^d}
\frac{(\tilde y^{(1)}_{\bx+\by,\bx})^2} {(\tildep{\bx+\by}+\tildep{\bx})^3}. 
\ee 
Here, energies are expressed in units of $\mu$ and momenta in units of $1/\xi$, i.e. $\tildep{\bx} = \ep{\bx/\xi}/\mu
=|x|(2+x^2)^{1/2}$, etc.

\begin{figure}\centering
\includegraphics[width=\linewidth]{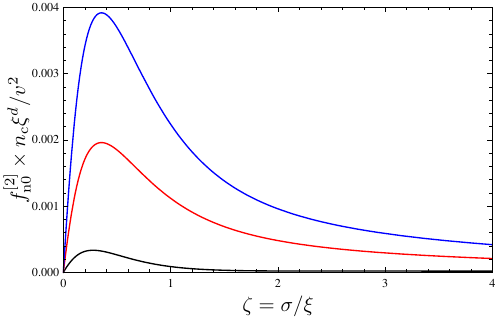}
\caption{Temperature-independent pair-bogolon contribution to the normal fraction, Eq.~\eqref{fnzero2.eq}, per
  potential variance $v^2=V^2/\mu^2$ as function of reduced potential correlation length $\zeta = \sigma/\xi$ in dimensions $d=1,2,3$ (top to bottom). }
\label{fn20.fig}
\end{figure}

At given potential strength $v^2=V^2/\mu^2$ and density $n\xi^d$,
the normal fraction \eqref{fnzero2.eq} is only a function of the reduced
correlation length $\zeta=\sigma/\xi$, plotted in Fig.~\ref{fn20.fig}.  
Just as the single-bogolon contribution of 
Fig.~\ref{fn1.fig}, also  this superfluid depletion is stronger in
lower dimensions. 
However, its dependence on potential
correlations is non-monotonic: the largest contribution is reached around
$\zeta\approx 0.3$, and it vanishes for both, very short- and
long-range correlated potentials.  

We can understand the asymptotic behavior for very small or large
correlation lengths by inspecting \eqref{fnzero2.eq} and
\eqref{Ndkernel}. In the Thomas-Fermi limit
$\zeta=\sigma/\xi\to\infty$ of a very smooth potential, the
potential correlator tends to the Dirac distribution,
$\zeta^dC_d(\zeta \by)\to(2\pi)^d\delta(\by)$.  In $d=2,3$, the kernel
$R_d(y)\to R_d(0)$ tends to a finite value, and the normal fraction \eqref{fnzero2.eq}
vanishes as $\zeta^{-2}$ because of the factor
$y_z^2$ under the integral. In $d=1$ the kernel diverges, but
only logarithmically, as $y\to 0$, implying that \eqref{fnzero2.eq} decreases slightly
more slowly, like $\zeta^{-2}\log\zeta$, which is not apparent on the scale of
Fig.~\ref{fn20.fig}.
 
In the opposite, white-noise regime $\zeta\to0$ the normal fraction \eqref{fnzero2.eq} vanishes linearly, in $d=3$ with the slope 
\be\label{fnzero2zetasmall}
\fn^{[2]}(0) = \frac{v^2}{n\xi^3} \frac{\zeta}{24(2\pi)^{3/2}} +O(\zeta^2),
\ee
with a similar behavior in $d=1,2$ (see
Fig.~\ref{fn20.fig}). Under the usual white-noise scaling $\zeta\to
0$, $v\to \infty$ at $v^2\zeta^d=const$, the depletion \eqref{fnzero2zetasmall} 
diverges as $\zeta^{1-d}$ in $d>1$ and should appear as a UV divergence in
treatments that assume a white-noise disorder from the start 
\cite{Giorgini1994,Lopatin2002}.  
In the present approach, keeping track of
finite potential correlations leads to finite
results under all circumstances.    

As evident from Fig.~\ref{fn20.fig}, even at its peak the temperature-independent pair-bogolon superfluid
depletion is more than three orders of magnitude smaller than the
single-bogolon value \eqref{fn1.eq}, all the more since it is 
multiplied by $1/n\xi^d\ll1$ (in $d=3$, this number translates to the gas parameter
$(na^3)^{1/2}$, and in $d=2$ to the interaction parameter $m g_2$),  
which must be a small for the Bogoliubov expansion to be
valid in the first place. 
This temperature-independent correction therefore can be safely neglected.

\bibliography{../sf}

\bibliographystyle{iopart-num}

\end{document}

%% file: settings.tex
\usepackage
{hyperref}	
\usepackage{amsmath,amsfonts,amssymb}

\usepackage
{graphicx}

\usepackage{xcolor}

\newcommand{\be}{\begin{equation}}
\newcommand{\ee}{\end{equation}}

\newcommand{\nc}{n_\text{c}}
\newcommand{\Nc}{N_\text{c}}

\newcommand{\vc}[1]{\mathbf{#1}}
\newcommand{\bg}{{\vc g}}
\newcommand{\br}{{\vc r}}
\newcommand{\bk}{{\vc k}}

\newcommand{\bq}{{\vc q}}
\newcommand{\bp}{{\vc p}}

\newcommand{\bx}{{\vc x}}
\newcommand{\by}{{\vc y}}

\newcommand{\eps}{\varepsilon} 

\newcommand{\calG}{\mathcal{G}}
\newcommand{\calV}{\mathcal{V}}
\newcommand{\rmd}{\mathrm{d}}
\newcommand{\intddr}{\int \rmd^d r}			

\newcommand{\dn}{\delta\hat{n}}			
\newcommand{\dph}{\delta\hat\varphi}		
\newcommand{\g}[1]{\hat\gamma_{#1}}		
\newcommand{\gd}[1]{\hat\gamma^\dagger_{#1}}	%

\newcommand{\epn}[1]{\eps^0_{#1}}		
\newcommand{\ep}[1]{\eps_{#1}}		
\newcommand{\tildep}[1]{\tilde{\eps}_{#1}}		

\newcommand{\xpct}[1]{\bigl\langle #1 \bigr\rangle} 
\newcommand{\avg}[1]{\overline{{#1}}}	

\newenvironment{smallpmatrix}{\left(\begin{smallmatrix}}%
{\end{smallmatrix}\right)}
 
\newcommand{\matr}[4]{\begin{pmatrix}
#1 & #2 \\
#3 & #4 
\end{pmatrix}
} 

\newcommand{\cvect}[2]{%
\begin{pmatrix} #1 \\ #2 \end{pmatrix}}

\newcommand{\rhos}{\rho_\text{s}}
\newcommand{\rhon}{\rho_\text{n}}
\newcommand{\rhonTF}{\rho_\text{nTF}}

\newcommand{\fn}{f_\text{n}}
\newcommand{\fnTF}[1]{f_\text{nTF}^{[#1]}}
\newcommand{\fnTFempty}{f_\text{nTF}}

\newcommand{\rhoc}{\rho_\text{c}}

\newcommand{\kB}{k_\text{B}}

\newcommand{\wtV}{\widetilde{V}}